\documentclass[twocolumn,showpacs,10pt]{revtex4-1}
\usepackage[colorlinks,citecolor=blue,linkcolor=red,dvipdfm]{hyperref}
\usepackage{amsmath,amssymb,graphicx}

\begin{document}

\title{One- and two-dimensional solitons under the action of the inverted cubic-quintic nonlinearity}
\author{Liangwei Zeng$^{1,2}$}
\author{Boris A. Malomed$^{3,4}$}
\author{Dumitru Mihalache$^5$}
\author{Xing Zhu$^1$}
\email{\underline{xingzhu@gzmtu.edu.cn}}

\affiliation{$^1$School of Arts and Sciences, Guangzhou Maritime University, Guangzhou 510725, China}

\affiliation{$^2$College of Physics and Optoelectronic Engineering, Shenzhen University, Shenzhen 518060, China}

\affiliation{$^3$Department of Physical Electronics, School of Electrical Engineering, Faculty of Engineering, and Center for Light-Matter Interaction, Tel Aviv University, P.O.B. 39040, Tel Aviv, Israel}

\affiliation{$^4$Instituto de Alta Investigaci\'{o}n, Universidad de Tarapac\'{a}, Casilla 7D, Arica, Chile}

\affiliation{$^5$Horia Hulubei National Institute of Physics and Nuclear Engineering, 077125 Magurele, Bucharest, Romania}

\begin{abstract}
The usual cubic-quintic (CQ) nonlinearity is proved to sustain one- and two-dimensional (1D and 2D) broad (flat-top) solitons. In this work, we demonstrate that 1D and 2D soliton families can be supported, in the semi-infinite bandgap (SIBG), by the interplay of a lattice potential and the nonlinearity including self-defocusing cubic and self-focusing quintic terms, with the sign combination inverted with respect to the usual CQ nonlinearity. The families include fundamental and dipole solitons in 1D, and fundamental, quadrupole, and vortex solitons in 2D. The power, shapes, and stability of the solitons are reported. The results are strongly affected by the positions of the solitons in SIBG, the families being unstable very close to or very far from the SIBG's edge. The inverted CQ nonlinearity, considered in this work, sustains sharp 1D and 2D stable solitons, which can be naturally used as bit pixels in photonic data-processing applications.
\end{abstract}

\maketitle

\section*{Introduction}

The formation of solitons \cite{REV1,REV2,REV3} is a fundamental topic in nonlinear physics \cite{REV4,REV5,REV6,REV7,HYD1}, especially in the fields of nonlinear optics and photonics \cite{OPT1,OPT2,OPT3,OPT4,OPT5,OPT6,OPT7} and quantum matter, such as Bose-Einstein condensates (BECs) \cite{BECa,BECb,BECc,BECd,BEC2,BEC3,BEC5}. Various types of soliton families have been reported, a majority of them representing one-dimensional (1D) states \cite{1DS2,1DS3,1DS4}. The creation of 2D solitons is a challenging issue, as self-focusing cubic and quintic nonlinear terms acting in the free 2D space, modelled by equations of the nonlinear-Schr\"{o}dinger (NLS) type, give rise to the critical and supercritical collapse, respectively, which makes all free-space solitons, supported by the self-focusing, unstable in 2D \cite{Berge,CLP3,CLP4,book}. To stabilize 2D and 3D solitons against the collapse, it was proposed to use linear and nonlinear potentials \cite{2DS2,YangMuss,2DS7,2DS8,book}. In particular, spatially periodic linear and nonlinear potentials, alias linear \cite{2DS2,YangMuss,LL1,LL2,LL3} and nonlinear \cite{NL1,NL2,NL3,NL4,LL4} lattices, can be used to build various types of stable 2D solitons, including fundamental ones \cite{SLL1}, dipoles \cite{SLLD1,SLLD2,SLLD3}, multipoles \cite{SLL2,SDF4}, solitary vortices \cite{SLL3,SLL3b}, and half-vortices \cite{SLL4}. The stable 2D solitons, pinned to the underlying lattice, offer a significant potential for the use as bit pixels in various data-processing schemes \cite{pixels1,pixels3}. Obviously, narrow 2D solitons are required to realize this application.

Another setting that provides stabilization of 2D \cite{MQT,Pego,CQ2,CQ4} and 3D \cite{3D,CQ5} solitons, including ones with embedded vorticity, includes competition of the self-focusing cubic and defocusing quintic nonlinear terms, which can be readily implemented in optical waveguides. In particular, stable 2D \cite{MQT,Pego} and 3D \cite{3D} vortex solitons, as well as the fundamental (zero-vorticity) ones \cite{Berntson}, supported by the cubic-quintic (CQ) nonlinearity have been predicted. A characteristic
feature of these modes is their flat-top shape, as the increase of the input optical power can be accommodated by the spatial expansion of the solitons, while their local intensity is bounded by the balance of the cubic self-focusing and quintic self-defocusing nonlinearities.

While the composite nonlinearity of this type is quite natural, as it appears as an approximate form of the saturable nonlinear response of the dielectric medium to the propagating electromagnetic waves, other types are physically relevant too. As demonstrated experimentally and explained theoretically \cite{Cid1,Cid2,Cid3}, the CQ nonlinearity (and its extension including the septimal term) \cite{CQSP1,CQSP2} can be efficiently engineered, including a possibility to separately choose the signs and
magnitudes of the cubic and quintic terms, in optical materials based on colloidal suspensions of metallic nanoparticles, using their radius, which takes values in the range of $1-100$ nm, and volume fraction $f$ of the nanoparticles, varying in the range of $10^{-5}-10^{-4}$, as control parameters. The effective nonlinearity is produced by the nanoparticles through the surface-plasmon-resonance mechanism.

The freedom in the engineering of the composite nonlinearities suggests one to consider the ``unusual" case of the \textit{inverted} CQ nonlinearity, composed of defocusing cubic and focusing quintic terms, which is the subject of the present work. The inverted CQ nonlinearity should be considered in the combination with a lattice potential, as, otherwise, there is no chance to construct any stable self-trapped state in the model. In this work, we demonstrate that this setting is promising for the above-mentioned applications, as the stable solitons can be maintained by the interplay of the cubic self-defocusing and quintic self-focusing nonlinearities in the desired form of narrow pixels, while the usual combination of the cubic focusing and quintic defocusing nonlinearities gives rise to the above-mentioned broad (flat-top) modes, which cannot be used as pixels.

It is relevant to mention that, in the case of the cubic or quintic self-defocusing per se, self-trapped modes can be found as gap solitons populating finite bandgaps induced by the lattice potential, while no bright solitons exist in the semi-infinite bandgap (SIBG) \cite{Salerno}. In the case of the self-focusing nonlinearity, solitons populate the SIBG, but do not exist in finite bandgaps. In the latter case, 2D solitons are unstable in the free-space SIBG (no lattice potential) due to the occurrence of the collapse \cite{Berge,CLP3,CLP4,book}. Nevertheless, both fundamental and vortical 2D solitons can be readily stabilized in the SIBG by the lattice potential in the self-focusing cubic medium \cite{2DS2,YangMuss}. The objective of the present work is to produce families of stable 1D and 2D bright solitons in SIBG under the combined action of the lattice potential and \textit{inverted} CQ nonlinearity. These are narrow (pixel-like) solitons of the fundamental and dipole types in 1D, and ones of the fundamental, quadrupole, and vortex types in 2D. The soliton solutions are constructed in the numerical form, and their stability is
identified by means of systematically performed simulations of the perturbed propagation. We conclude that the position of the solitons in the SIBG essentially affects their shape and stability.

The subsequent presentation is arranged as follows. Systematically collected numerical results for the soliton families are presented in the section of results, which is divided in two parts, reporting the findings for 1D and 2D solitons. The paper is concluded by the section of conclusion. Then the model is introduced in section of methods, where we also produce some simple analytical results for broad and narrow solitons, which suggest their stability in terms of the well-known Vakhitov-Kolokolov (VK) criterion \cite{VK,Berge}.

\begin{figure}[tbp]
\begin{center}
\includegraphics[width=1\columnwidth]{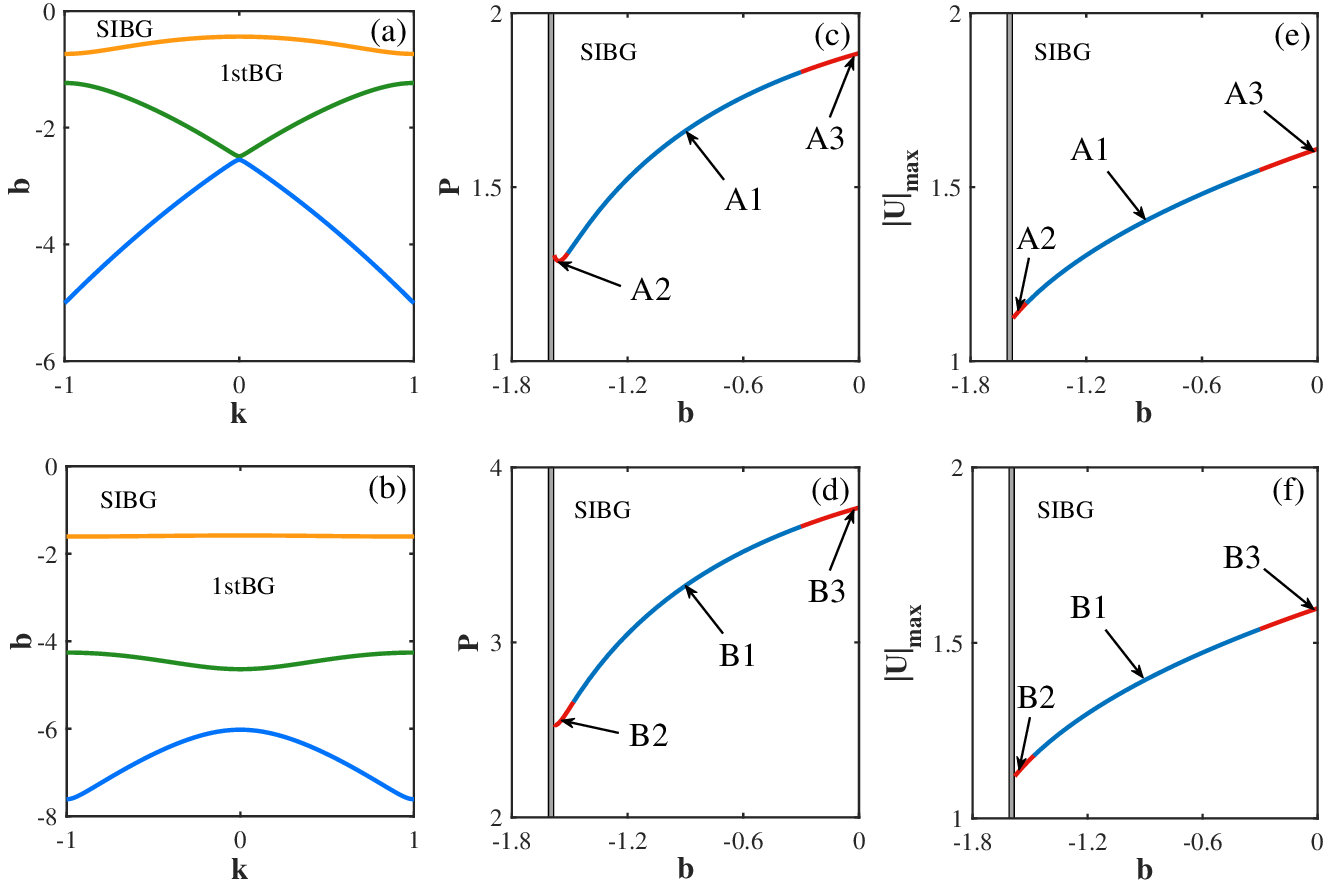}
\end{center}
\caption{\textbf{Bandgap spectra, power and amplitude of 1D soliton families.} (a,b): The bandgap spectrum produced by the 1D version of the linearized equation (\protect\ref{NLSES}) with potential (\protect\ref{V1D}), for $V_{0}=1$ (a) and $V_{0}=6$ (b). The spectrum is plotted in the plane of the quasi-momentum $k$ of Bloch modes and propagation constant $b$. The red, blue, and green strips represent the first, second, and third Bloch bands, respectively. Acronyms SIBG and 1stBG stand for the semi-infinite and first bandgaps, respectively. (c,d): The soliton's power $P$ vs. propagation constant $b$ in the SIBG for families of 1D fundamental (c) and dipole (d) solitons in the deep lattice potential, with $V_{0}=6$. Blue and red segments of the curves represent stable and unstable solitons, respectively. (e,f): The amplitude (maximum value of $|U(x)|$) of the families of 1D fundamental (e) and dipole (f) solitons at $V_{0}=6$. The vertical grey areas in panels (c)--(f) stand for the first Bloch band.}
\label{fig1}
\end{figure}

\begin{figure}[tbp]
\begin{center}
\includegraphics[width=1\columnwidth]{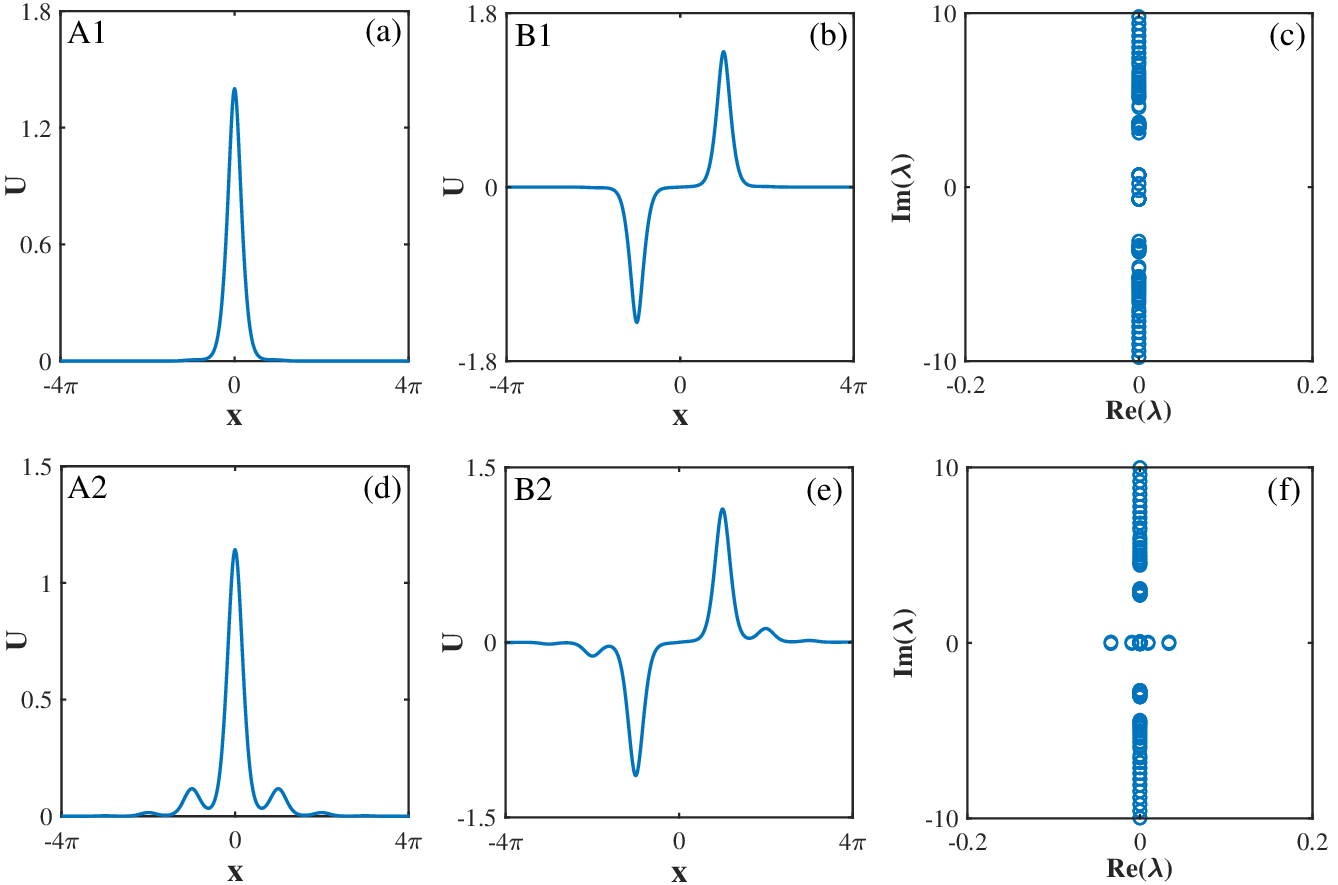}
\end{center}
\caption{\textbf{Profiles and eigenvalues of linear stability analysis for 1D soliton families.} Profiles of the 1D stable fundamental (a) and dipole (b) soliton found at $b=-0.9$, which correspond, respectively, to labels A1 and B1 in Figs. \protect\ref{fig1}(c,d). (c): Eigenvalues produced by the numerical solution of Eq. (\protect\ref{LSA}) for the soliton in panel (b). Unstable fundamental and dipole solitons, found at $b=-1.55$, which correspond to labels A2 and B2 in Figs. \protect\ref{fig1}(c,d), are plotted in panels (d)
and (e), respectively. (f): Eigenvalues $\protect\lambda $ produced by Eq. (\protect\ref{LSA}) for the soliton in panel (e). The corresponding solutions of Eq. (\protect\ref{NLSES}) are obtained for potential (\protect\ref{V1D}) with $V_{0}=6$.}
\label{fig2}
\end{figure}

\begin{figure*}[tbp]
\begin{center}
\includegraphics[width=1.8\columnwidth]{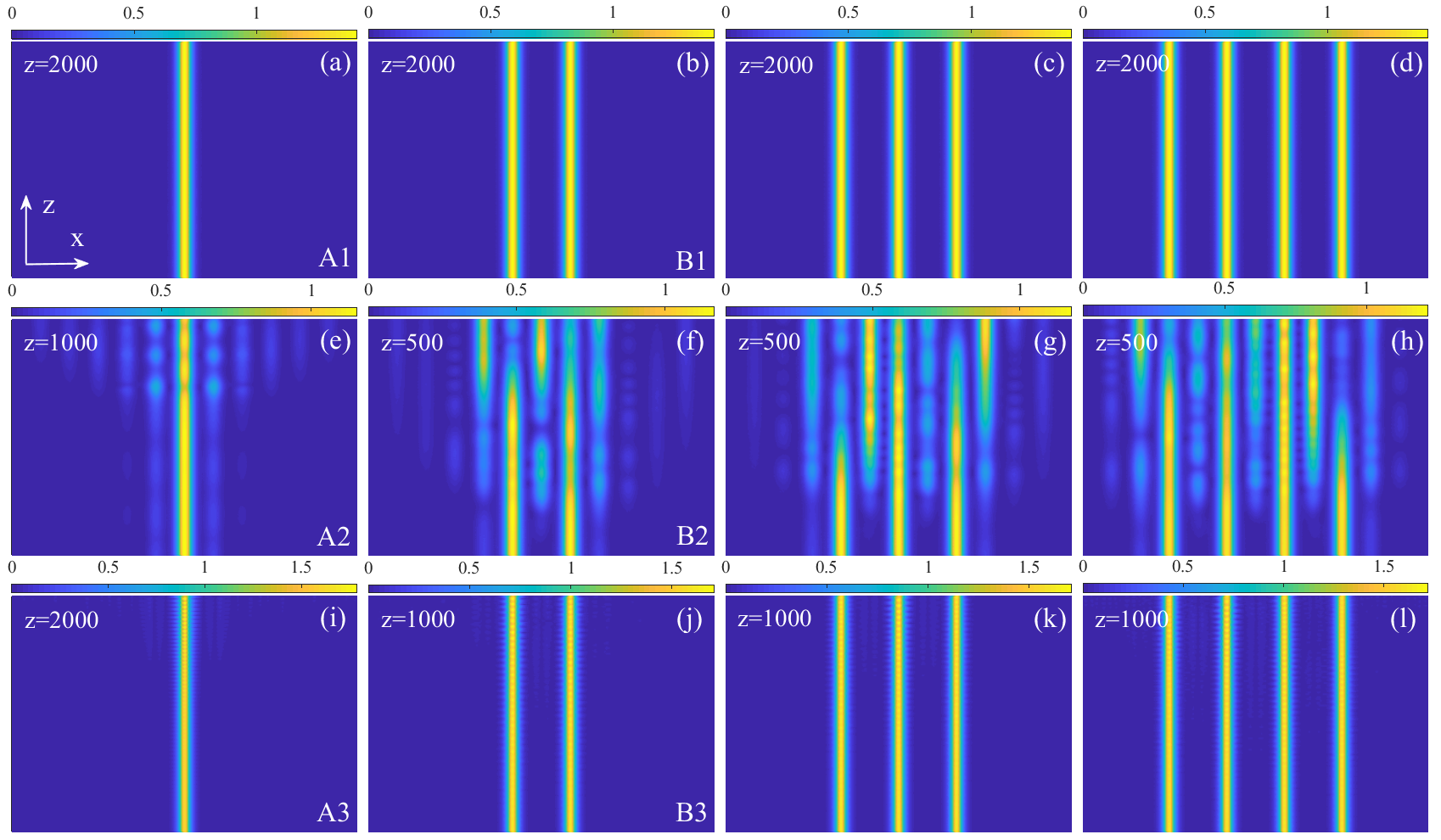}
\end{center}
\caption{\textbf{Perturbed propagations of 1D soliton families.} The top row displays the stable perturbed propagation of 1D solitons with $b=-0.9$: (a) the fundamental soliton (which corresponds to point A1 in Fig. \protect\ref{fig1}(c)); (b) the dipole soliton (corresponding to point B1 in Fig. \protect\ref{fig1}(d)); (c) the tripole soliton; (d) the quadrupole one. The middle row displays the unstable propagation of the solitons with $b=-1.55$: (e) the fundamental soliton (corresponding to point A2 in Fig. \protect\ref{fig1}(c); (f) the dipole soliton (corresponding to point B2 in Fig. \protect\ref{fig1}(d)); (g) the tripole soliton; (h) the quadrupole one. The bottom row displays the unstable propagation of the solitons with $b=-0.02$: (i) the fundamental soliton (corresponding to point A3 in Fig. \protect\ref{fig1}(c); (j) the dipole soliton (corresponding to point B3 in Fig. \protect\ref{fig1}(d)); (k) the tripole soliton; (l) the quadrupole one.}
\label{fig3}
\end{figure*}

\section*{Results}
\label{sec2}

\subsection*{One-dimensional solitons}
\label{sec2a}

The Bloch bandgap structure \cite{1DBG1,1DBG2} produced by the linearization of the 1D version of Eq. (\ref{NLSES}) with lattice potential (\ref{V1D}) is plotted in Figs. \ref{fig1}(a,b), for the moderate ($V_{0}=1$) and deep ($V_{0}=6)$ potentials, respectively, with SIBG and 1stBG standing for the semi-infinite bandgap and the first finite bandgap, respectively, which are separated by thin Bloch bands plotted by curves of different colors.

It is observed that solely the first finite bandgap is open at $V_{0}=1$, while there are two of them at $V_{0}=6$. Here we address 1D solitons in SIBG. The family of the fundamental solitons is represented in Fig. \ref{fig1}(c) by the respective dependence of the soliton's power $P$ (see Eq. (\ref{SP1})) on propagation constant $b$. The $P(b)$ features a narrow VK-unstable interval with $dP/db<0$, followed by a broad one with $dP/db>0$. In addition, a family of 1D dipole solitons is represented by the corresponding $P(b)$ curve in Fig. \ref{fig1}(d). Naturally, for any given $b$, the power of the dipoles is approximately twice its counterparts for the fundamental solitons plotted in Fig. \ref{fig1}(c).

Blue and red segments of the curves displayed in Figs. \ref{fig1}(c,d) represent stable and unstable parts of the respective soliton families. Obviously, all stable subfamilies satisfy the VK criterion, $dP/db>0$,
which, as said above, is a necessary (but not sufficient) condition for the stability of solitons \cite{VK,Berge}. The transition to instability of the fundamental and dipole solitons at larger values of the power (at $P>$ $1.83$ and $P>3.66$, respectively, which correspond, approximately, to $-b<0.3$), i.e., deeper in the SIBG, is a natural effect, as, deeply enough, the effect of the lattice potential becomes immaterial, and the combination of the defocusing cubic and focusing quintic nonlinear terms leads to
instability as usual.

Profiles of typical 1D fundamental and dipole solitons, which are marked by labels A1, A2 and B1, B2 in Figs. \ref{fig1}(c,d) are plotted in Fig. \ref{fig2}. It is observed that, in accordance with the above-mentioned expectation, the solitons take the shape of narrow pixels, which makes them appropriate for applications. Further, note that each peak of $|U(x)|$ of the stable dipole, plotted at $b=-0.9$ in Fig. \ref{fig2}(b), is similar to the stable fundamental soliton, which is plotted in Fig. \ref{fig2}(a) for
the same value of $b$. While the stable solitons, such the fundamental and dipole ones, displayed here for $b=-0.9$, which reside deep in the SIBG, feature, respectively, the simple single- or double-peak shapes, unstable solitons, which reside close to the SIBG edge (such as the ones displayed in Figs. \ref{fig2}(d,e) for $b=-1.55$), demonstrate additional lower peaks near the main ones, which makes their shape essentially different from that of pixels. This feature is naturally explained by the fact that the unstable solitons are located close to the Bloch modes, which are represented by spatially periodic multi-peak patterns. The eigenvalues $\lambda $, produced by the numerical solution of Eq. (\ref{LSA}) for the solitons in panels (b)
and (e), are presented in panels (c) and (f), respectively. Obviously, panel (c) implies that the soliton in panel (b) is stable, while the one in panel (e) is unstable, according to panel (f).

To further investigate the variation of the solitons with the decrease of $|b|$ for the 1D solitons, we display their amplitude $|U(x)|_{\max }$ vs. the propagation constant $b$ for 1D fundamental and dipole gap solitons in Figs. \ref{fig1}(e,f), respectively. It is obvious that the amplitude grows monotonously with the decrease of $|b|$, being nearly identical for the fundamental solitons and dipoles.

The (in)stability of the 1D solitons, which are represented by the blue (stable) and red (unstable) colors of the $P(b)$ curves in Figs. \ref{fig1}(c,d), was corroborated by simulations of the perturbed evolution, examples of which are presented in Fig. \ref{fig3}, including the evolution of the solitons whose stationary shape is displayed in Fig. \ref{fig2}, which correspond to $b=-0.9$ and $b=-1.55$ for stable and unstable ones, respectively. In particular, in this and other cases, the unstable solitons exhibit gradual decay in the course of the propagation.

In addition to the fundamental and dipole solitons, Fig. \ref{fig3} also exhibits examples of stable and unstable higher-order solitons, \textit{viz}., tripole and quadrupole ones, which can be readily constructed as additional solutions of Eq. (\ref{NLSES}). As well as the fundamental and dipole solitons, the higher-order ones are stable for $b=-0.9$ and unstable for $b=-1.55$.

The evolution of unstable fundamental and dipole solitons, which belong to the red high-power segments in Figs. \ref{fig1}(c,d), is displayed in panels (a) and (j) of Fig. \ref{fig3}. The higher-order unstable solitons with high power are presented there too, featuring weak oscillations in the course of the evolution.

\begin{figure}[tbp]
\begin{center}
\includegraphics[width=1\columnwidth]{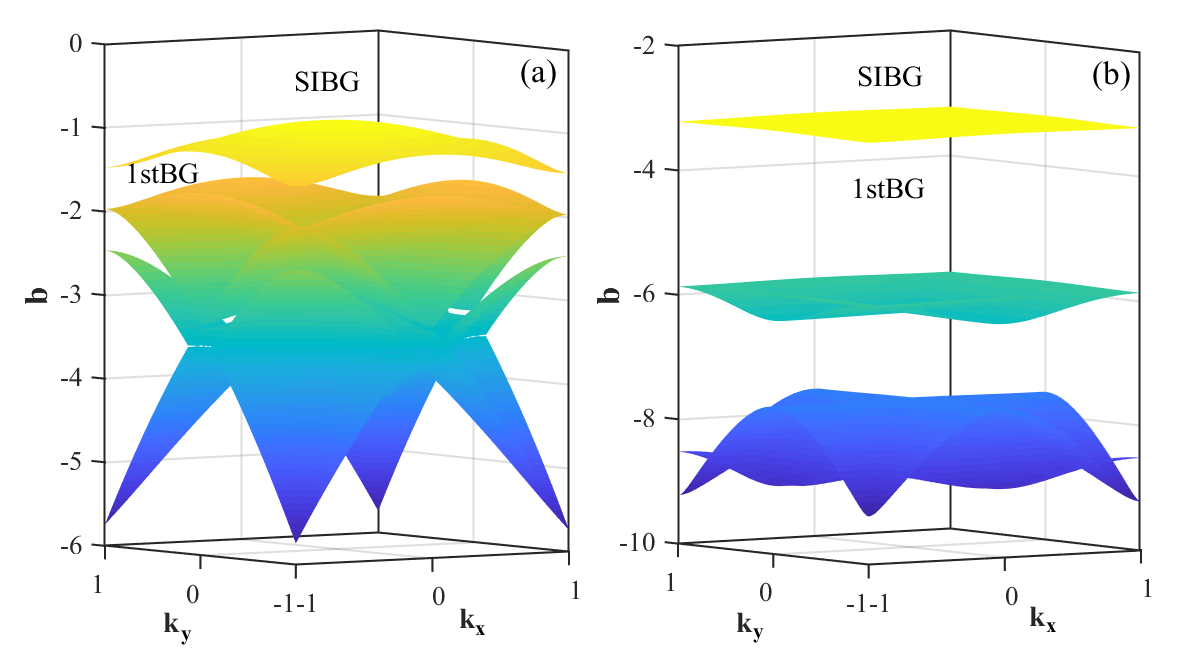}
\end{center}
\caption{\textbf{The 2D Bandgap spectra.} (a,b): The bandgap spectrum produced by the linear version of the
2D equation (\protect\ref{NLSES}) with potential (\protect\ref{V2D}), for $V_{0}=1$ (a) and $V_{0}=6$ (b). The spectrum shows the propagation constant $b$ of Bloch modes vs. the components $k_{x}$ and $k_{y}$ of their quasi-momentum. As in Fig. \protect\ref{fig1}, acronyms SIBG and 1stBG stand for the semi-infinite and the first bandgaps, respectively. In panel (a), surfaces denote, from top tp bottom, the first, second, third, fourth, and fifth Bloch bands.}
\label{fig4}
\end{figure}

\begin{figure}[tbp]
\begin{center}
\includegraphics[width=1\columnwidth]{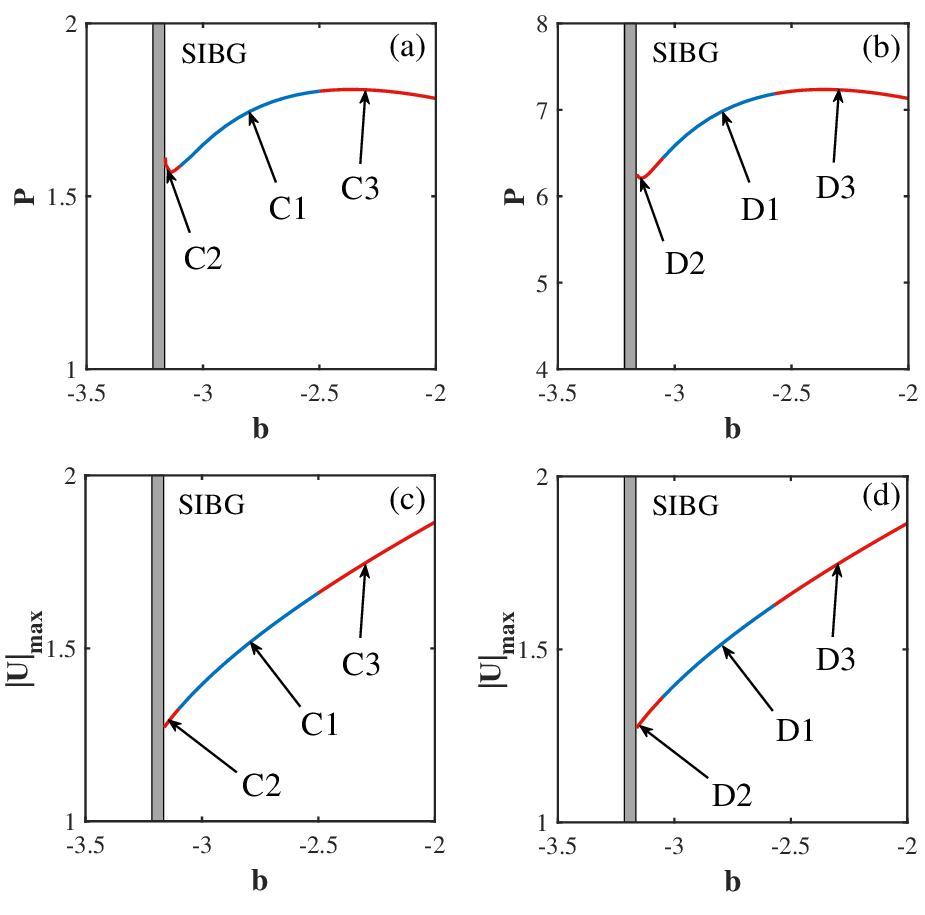}
\end{center}
\caption{\textbf{Power and amplitude of 2D soliton families.} The curves of the soliton power $P$ versus the propagation constant $b$ in the semi-infinite band gap of the 2D soliton families at $V_{0}=6$: (a) for the fundamental solitons; (b) for the quadrupoles. The amplitude (maximum value of $|U(x,y)|$) of the 2D soliton families at $V_{0}=6$: (c) for the fundamental solitons; (d) for the quadrupoles. The vertical grey areas in panels (a)--(d) stand for the first Bloch band.}
\label{fig5}
\end{figure}

\subsection*{Two-dimensional solitons}
\label{sec2b}

The Bloch bandgap spectrum \cite{2DBG1,2DBG2,2DBG3} produced by the linearized version of the 2D equation (\ref{NLSES}) is presented in Fig. \ref{fig4} for the 2D potential (\ref{V2D}) with $V_{0}=1$ and $V_{0}=6$. As in Fig. \ref{fig1}, SIBG and 1stBG denote the semi-infinite and first finite bandgaps, respectively. Panel (a) shows that only a narrow first bandgap opens with $V_{0}=1$, while both the first and second ones open with $V_{0}=6 $ in panel (b). As above, we here focus on solitons populating SIBG.

Families of 2D fundamental and quadrupole solitons, produced by the numerical solution of Eq. (\ref{NLSES}), are represented by the corresponding $P(b)$ curves (for the 2D integral power defined as per Eq. (\ref{SP2})) in Figs. \ref{fig5}(a) and (b), where, similar to Figs. \ref{fig1}(c,d), stable and unstable subfamilies are designated by the blue and red colors, respectively. Naturally, for any given $b$ the total power (\ref{SP2}) of the quadrupole soliton is almost exactly fourfold the power of the fundamental soliton with the same $b$. A difference from the similar results for families of fundamental and dipole solitons in 1D, presented above in Figs. \ref{fig1}(c,d), is that the transition to the instability deep inside SIBG is more pronounced (which is natural in the 2D case) and is explicitly related to the breakup of the VK criterion.

\begin{figure}[tbp]
\begin{center}
\includegraphics[width=1\columnwidth]{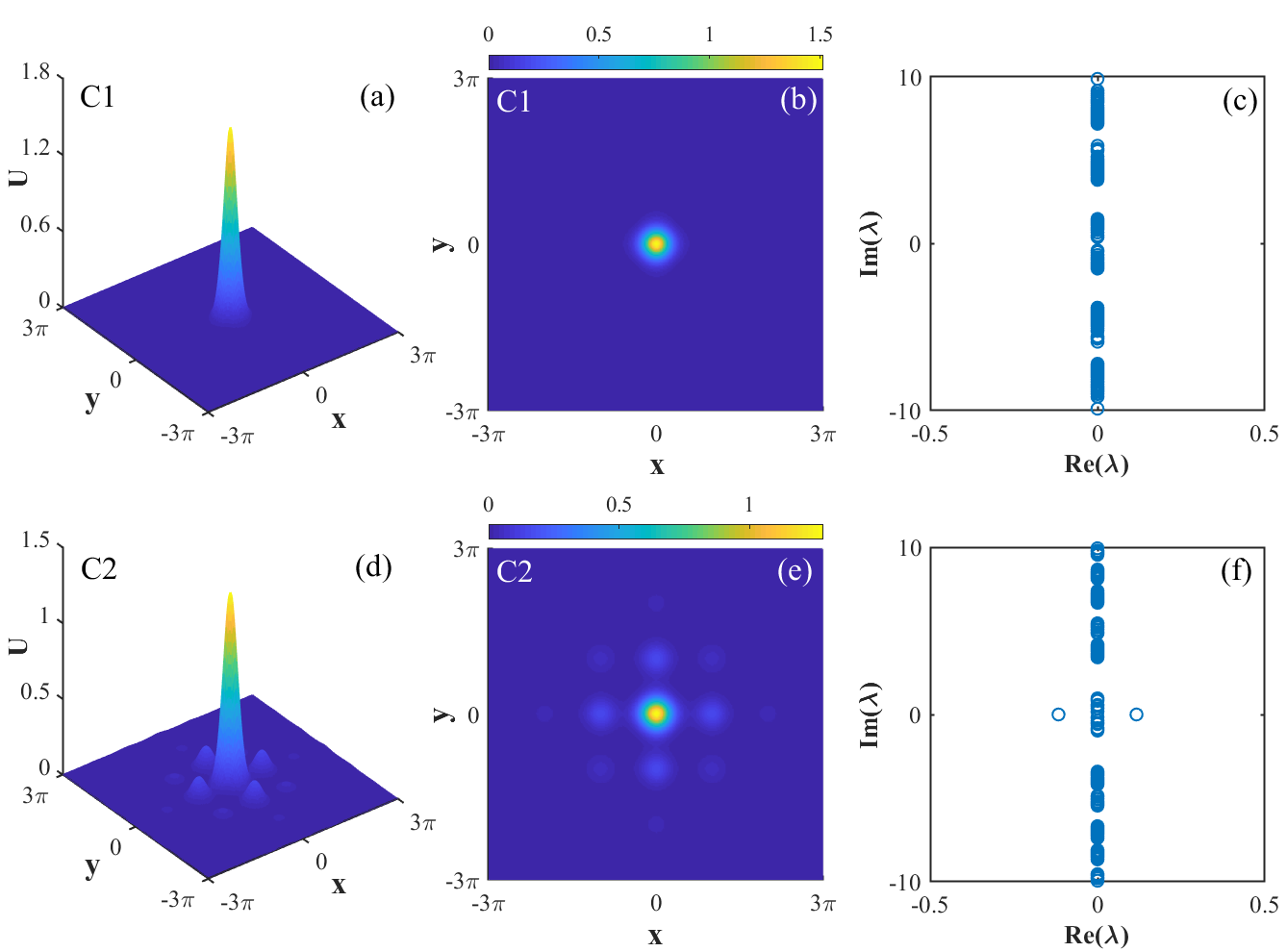}
\end{center}
\caption{\textbf{The 3D views, contour maps and eigenvalues of linear stability analysis for 2D fundamental solitons.} The 3D view (a), contour map of $|U\left( x,y\right) |$ (b), and eigenvalues $\protect\lambda $ produced by Eq. (\protect\ref{LSA}) (c) for the stable 2D fundamental soliton, labeled C1 in Fig. \protect\ref{fig5}(a), which is obtained as the numerical solution of Eq. (\protect\ref{NLSES}) with $b=-2.8$ and depth $V_{0}=6$ of the lattice potential (\protect\ref{V2D}). Panels (d)--(f) show the same, but for the unstable soliton with $b=-3.15$, which is labeled C2 in Fig. \protect\ref{fig5}(a).}
\label{fig6}
\end{figure}

\begin{figure}[tbp]
\begin{center}
\includegraphics[width=1\columnwidth]{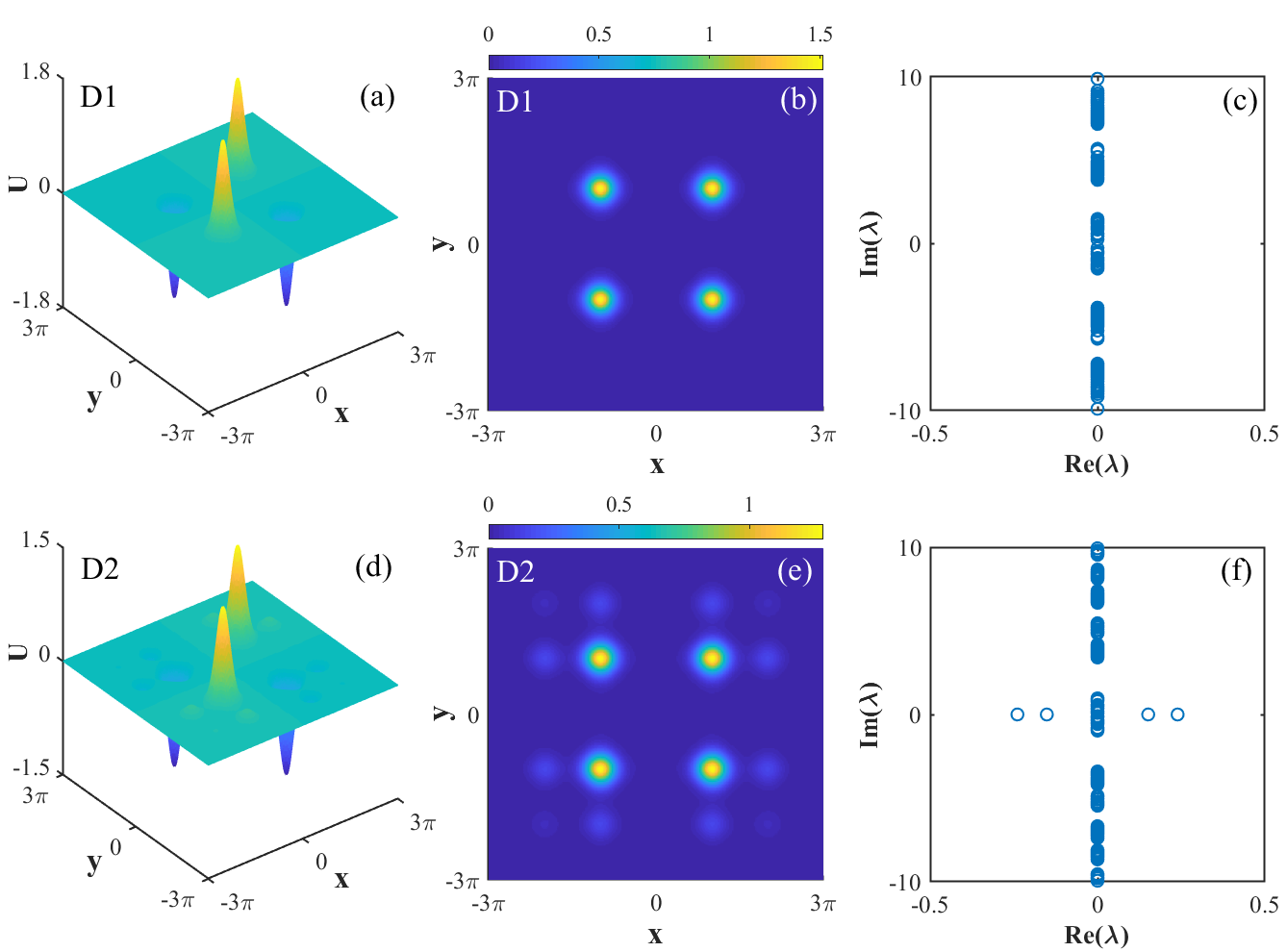}
\end{center}
\caption{\textbf{The 3D views, contour maps and eigenvalues of linear stability analysis for 2D quadrupole solitons.} The 3D view (a), contour map of $|U\left( x,y\right) |$ (b), and eigenvalues $\protect\lambda $ produced by Eq. (\protect\ref{LSA}) (c) for the stable 2D quadrupole soliton, labeled D1 in Fig. \protect\ref{fig5}(b), which is obtained as the numerical solution of Eq. (\protect\ref{NLSES}) with $b=-2.8$ and depth $V_{0}=6$ of the lattice potential (\protect\ref{V2D}). Panels (d)--(f) show the same, but for the unstable soliton with $b=-3.15$, which is labeled D2 in Fig. \protect\ref{fig5}(b).}
\label{fig7}
\end{figure}

Similar to the situation reported for the 1D solitons in Figs. \ref{fig1}(c,d), stable subfamilies of the 2D solitons obey the VK criterion in Fig. \ref{fig5}(a,b). On the other hand, an essential difference from the results for the 1D model is that the $P(b)$ curves for the 2D solitons include relatively broad unstable segments deeper inside SIBG (specifically, these are ones at $-b<2.5$ for the fundamental solitons, and at $-b<2.57$ for quadrupoles), whose instability (unlike that of the above-mentioned narrow intervals of 1D unstable solitons in Fig. \ref{fig1}(c,d), at $-b<0.3$) is directly explained by the violation of the VK criterion. Indeed, the strongly destabilizing effect of the quintic self-focusing term deeply in SIBG, where the effect of the lattice potential is immaterial, is a natural feature in the 2D setting.

The families of the 2D fundamental and quadrupole solitons are additionally characterized, in Figs. \ref{fig5}(c) and (d), by the respective dependences of their amplitude, $|U|_{\max }$, on the propagation constant $b$. These dependences, which are nearly identical for the fundamental solitons and quadrupoles, are quite similar to their counterparts for the 1D fundamental and dipole solitons, cf. Figs. \ref{fig1}(e,f).

\begin{figure}[tbp]
\begin{center}
\includegraphics[width=1\columnwidth]{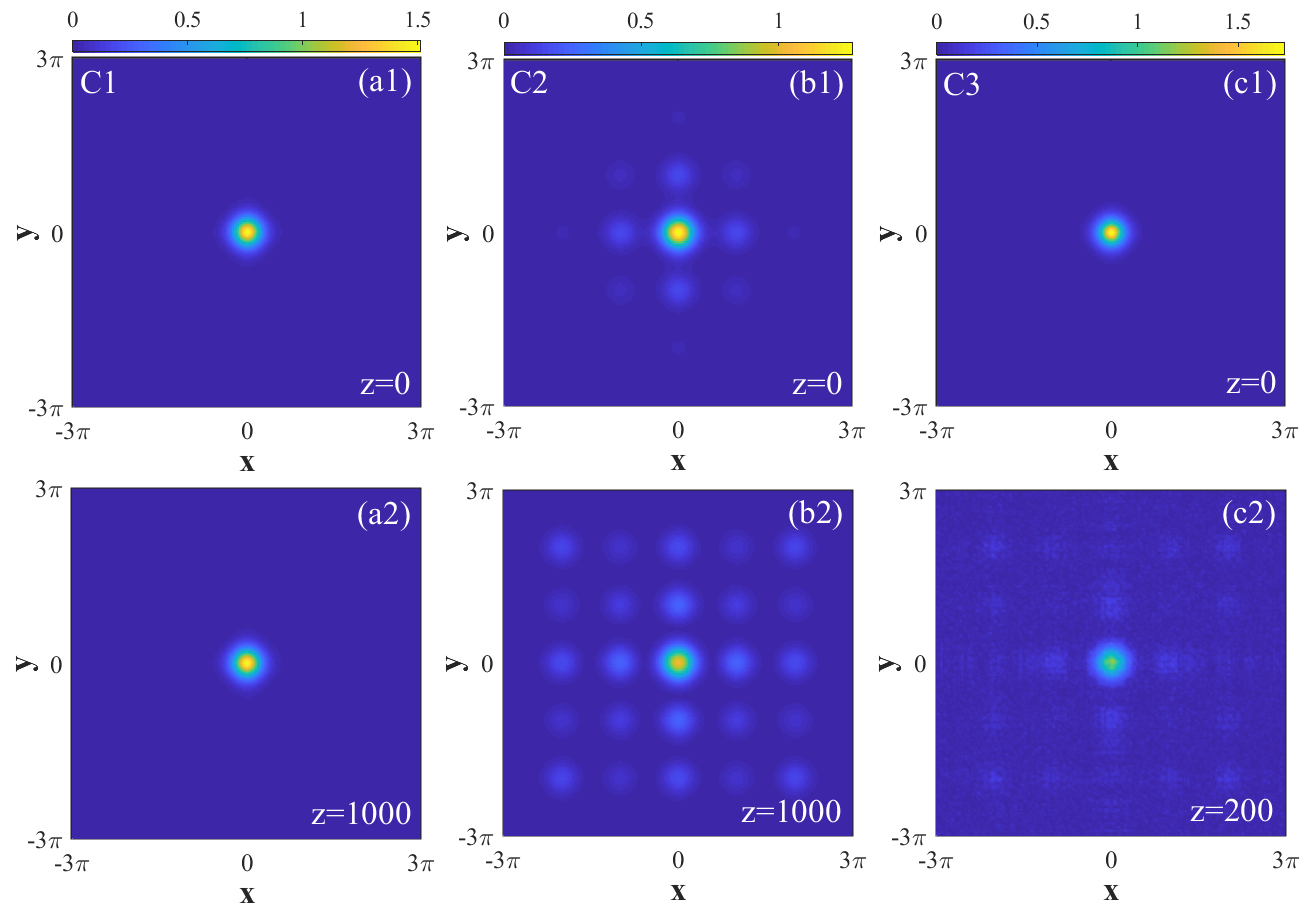}
\end{center}
\caption{\textbf{Perturbed propagations of 2D fundamental solitons.} The perturbed propagation of 2D fundamental solitons in the framework of Eq. (\protect\ref{NLSE}) with depth $V_{0}=6$ of potential (\protect\ref{V2D}): (a1)--(a2) the stable propagation with $b=-2.8$, which corresponds to label C1 in Fig. \protect\ref{fig5}(a); (b1)--(b2): the unstable propagation with $b=-3.15$, which corresponds to label C2 in Fig. \protect\ref{fig5}(a); (c1)--(c2): the unstable propagation with $b=-2.3$.}
\label{fig8}
\end{figure}

\begin{figure}[tbp]
\begin{center}
\includegraphics[width=1\columnwidth]{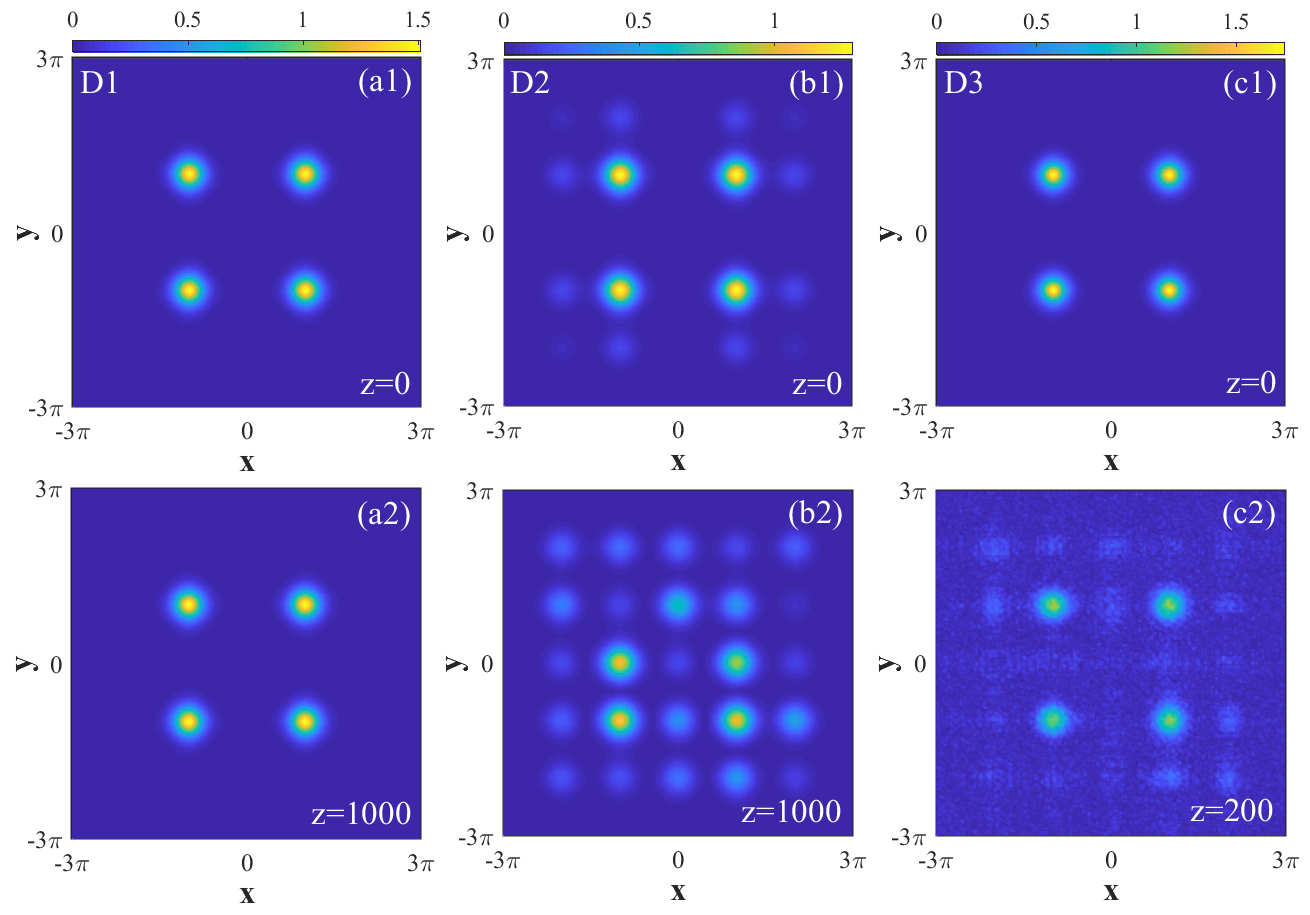}
\end{center}
\caption{\textbf{Perturbed propagations of 2D quadrupole solitons.} The perturbed propagation of 2D quadrupole solitons in the framework of Eq. (\protect\ref{NLSE}) with depth $V_{0}=6$ of potential (\protect\ref{V2D}): (a1)--(a2): the stable propagation for $b=-2.8$, which corresponds to label D1 in Fig. \protect\ref{fig5}(b); (b1)--(b2): the unstable propagation with $b=-3.15$, which corresponds to label D2 in Fig. \protect\ref{fig5}(b); (c1)--(c2): the unstable propagation with $b=-2.3$.}
\label{fig9}
\end{figure}

The shapes the 2D fundamental solitons and quadrupoles, labeled by C1, C2 and D1, D2 in Figs. \ref{fig5}(a) and (b), are plotted in Figs. \ref{fig6} and \ref{fig7}, respectively, by means of 3D views and power contour plots in the $\left(x,y\right) $ plane. Similar to what is reported above for the 1D solitons in
Fig. \ref{fig2}, the stable 2D solitons and quadrupoles, located relatively deep in SIBG, feature isolated sharp peaks (a single one for the fundamental soliton, and four identical ones for the quadrupole), thus corroborating their potential use as pixels in the applications, while the unstable solitons and quadrupoles, residing close to the SIBG edge, exhibit additional small peaks around the major ones, which makes them different from pixels. The eigenvalues $\lambda $ of the instability growth rate for these 2D modes are displayed in the right columns of Figs. \ref{fig6} and \ref{fig7}. In addition, the perturbed propagation of the fundamental and quadrupole solitons are displayed in Figs. \protect\ref{fig8} and \protect\ref{fig9}, respectively.

\begin{figure}[tbp]
\begin{center}
\includegraphics[width=1\columnwidth]{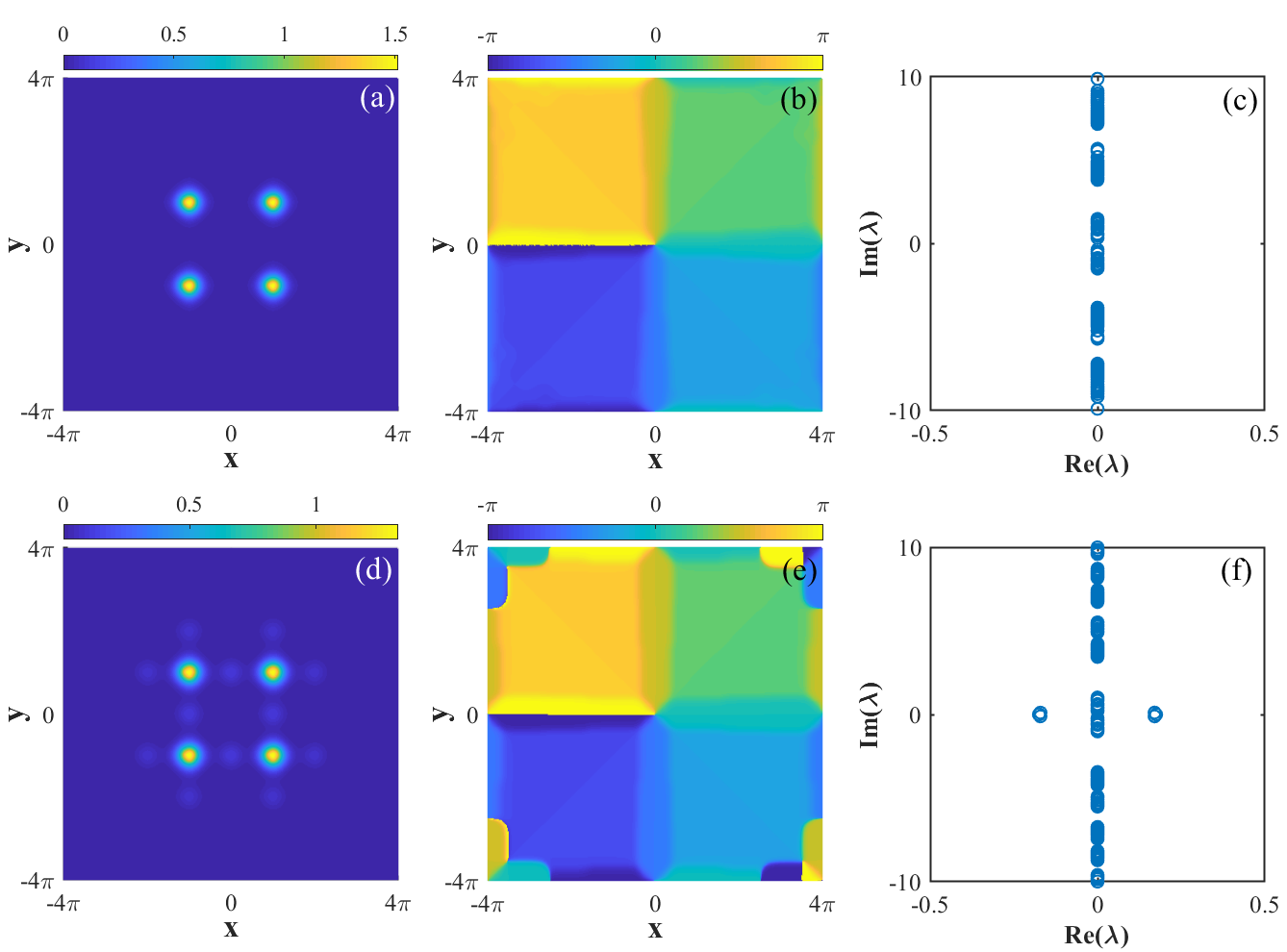}
\end{center}
\caption{\textbf{Contours, phases and eigenvalues of linear stability analysis for vortex solitons.} Contours, phases, and (in)stability eigenvalues $\protect\lambda $ for stable and unstable vortex solitons with $b=-2.8$ in \ (a)--(c), and $b=-3.1$ in (d)--(f), respectively.}
\label{fig10}
\end{figure}

\begin{figure}[tbp]
\begin{center}
\includegraphics[width=1\columnwidth]{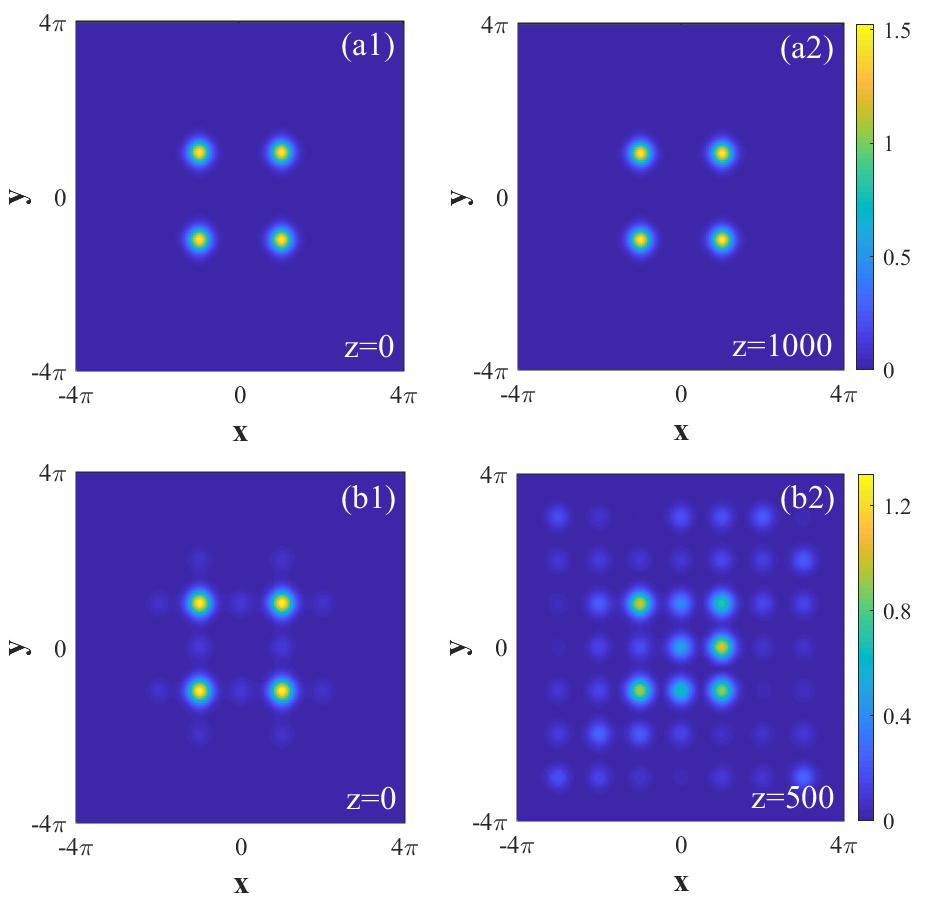}
\end{center}
\caption{\textbf{Perturbed propagations of vortex solitons.} The perturbed propagation of vortex solitons in the framework of Eq. (\protect\ref{NLSE}) with depth $V_{0}=6$ of potential (\protect\ref{V2D}): (a1)--(a2): the stable propagation of the vortex soliton at $b=-2.8$; (b1)--(b2): the unstable propagation of the vortex soliton at $b=-3.1$.}
\label{fig11}
\end{figure}

The stability of 2D solitons was also identified by means of systematic numerical simulations of their perturbed propagation. Typical examples of the stable and unstable propagation of the 2D fundamental soliton with propagation constants $b=-2.8$ and $-3.15$ are displayed in the left and middle columns of Fig. \ref{fig8}. In particular, the unstable mode suffers gradual decay in the course of the propagation, similar to the instability of the 1D solitons (cf. Figs. \ref{fig3}(e-h)). In addition, the unstable propagation of 2D fundamental solitons deep in SIBG (with $b=-2.3$) is displayed in the right column of Fig. \ref{fig8}. It is seen that the latter soliton suffers distortion in the course of the propagation.

Similar results for the perturbed propagation of stable and unstable quadrupole solitons are presented in Fig. \ref{fig9}. The gradual decay of unstable quadrupoles is similar to that exhibited by the unstable 2D
fundamental solitons.

Solitons with embedded vorticity are also supported in the present model, as shown in Fig. \ref{fig10} by means of the contour and phases plots, and the (in)stability eigenvalues for the vortex solitons with $b=-2.8$ and $-3.1 $. According to panel (c), the vortex from panel (a) is stable. On the other hand, the vortex soliton in panel (d) is unstable, according to (f).

The perturbed propagations of the vortex solitons from Fig. \ref{fig10} is displayed in Fig. \ref{fig11}. This figure corroborates their stability in panels (a1,a2) and instability in (b1,b2), respectively. It is seen that the stable vortex soliton keeps its integrity in the course of the long-distance propagation (the top row), while the unstable one is destructed (the bottom row).

\section*{Conclusion}

\label{sec3}

We have demonstrated that stable 1D and 2D solitons of several types (fundamental solitons, dipoles, quadrupoles, and vortices), belonging to the SIBG (semi-infinite bandgap) in the system's spectrum, can be sustained by the unusual (\textit{inverted}) but physically relevant combination of the self-defocusing cubic and focusing quintic nonlinearities, in the combination with the lattice potentials. On the contrary to the broad (flat-top) solitons supported by the usual CQ (cubic-quintic) nonlinearity, the inverted setting gives rise to stable narrow 1D and 2D ones, which may be used as bit pixels in photonic data-processing schemes. The inverted form of the CQ nonlinearity can be realized experimentally in terms of the light propagation in a colloidal material containing metallic nanoparticles. The soliton modes produced in this work are characterized by their shape, power, and stability, which are essentially affected by the position of the solitons in SIBG (semi-infinite bandgap). Stable 1D and 2D soliton families obey the well-known VK (Vakhitov-Kolokolov) stability criterion, the solitons being unstable in narrow intervals of the propagation constant near the SIBG's edge. Unlike the sharp (pixel-like) stable solitons, the unstable
ones feature profiles that include  low-amplitude peaks in addition to the sharp central ones. The solitons are also unstable deep inside SIBG, where the effect of the lattice potential becomes immaterial, and the combination of the defocusing cubic and focusing quintic terms naturally leads to the instability.

The model considered in this work may be realized in optical media. In particular, a natural implementation of the 1D and 2D settings is possible, respectively, in planar and bulk waveguides built in colloidal suspensions of metallic nanoparticles \cite{Cid1,Cid2,Cid3}. The effective lattice potentials can be induced by spatially patterned distributions of dopants in the waveguide, which affect the linear interaction of the propagating light with the underlying material, cf. Ref. \cite{Kip}.

The results reported in this work are helpful for the comprehensive understanding of the bright solitons supported by competing nonlinearities, such as that represented by the CQ terms with the inverted combination of their signs (defocusing cubic and focusing quintic). The solitons of the fundamental and higher-order types (dipoles, multipoles, and vortices) are considered in the 1D and 2D geometries. The stable solitons, featuring narrow shapes, may find applications as pixels in photonic setups.

As an extension of the work, it may be interesting to explore vortex solitons with higher values of the
topological charge, and 2D solitons in the model combining the inverted CQ nonlinearity in combination with lattice potentials of other types, such as triangular, hexagonal, and quasiperiodic. In addition, soliton families in two-component systems with the inverted CQ nonlinearity may also be an interesting subject, including the fundamental, dipole, multipole, and vortex solitons.

\section*{Methods}
\label{sec4}

\subsection*{The basic equations}

The propagation of amplitude $E\left( x,y;z\right)$ of the optical wave under the action of the inverted CQ nonlinearity, with the cubic and quintic coefficients, $\mathrm{g}>0$ and $\xi <0$, and 2D lattice potential $V\left( x,y\right)$, is governed by the respective NLS equation, written in the scaled form \cite{CQ4,NLSECQ}:
\begin{equation}
i\frac{\partial E}{\partial z}=-\frac{1}{2}\nabla ^{2}E+V\left( x,y\right) E+\mathrm{g}\left\vert E\right\vert ^{2}E+\xi \left\vert E\right\vert ^{4}E
\label{NLSE}
\end{equation}
(or its 1D reduction). Here, $z$ is the propagation distance, and the paraxial-diffraction operator, $\nabla ^{2}=\partial ^{2}/\partial x^{2}+\partial ^{2}/\partial y^{2}$, acts on the transverse coordinates, $\left( x,y\right) $. The lattice potentials with depth $2V_{0}>0$ (or $V_{0}>0$, in 1D) are taken in the usual form \cite{2DBG1},
\begin{equation}
V_{\mathrm{2D}}=V_{0}~(\mathrm{sin}^{2}x+\mathrm{sin}^{2}y),  \label{V2D}
\end{equation}
\begin{equation}
V_{\mathrm{1D}}=V_{0}~\mathrm{sin}^{2}x.  \label{V1D}
\end{equation}
The numerical results are reported below for coefficients $\mathrm{g}=1$ and $\xi =-1$ (one of these values is fixed by scaling, while the choice of the other one makes it possible to produce generic results).

Stationary solutions of Eq. (\ref{NLSE}) with a real propagation constant $b$ are looked for as
\begin{equation}
E=U\left( x,y\right) ~\mathrm{exp}(ibz),  \label{U}
\end{equation}
where the stationary wave function $U$ satisfies the equation
\begin{equation}
-bU=-\frac{1}{2}\nabla ^{2}U+V\left( x,y\right) U+\mathrm{g}|U|^{2}U+\xi
|U|^{4}U.  \label{NLSES}
\end{equation}
The 2D and 1D stationary solutions are characterized by their total power:
\newline
\begin{equation}
P_{\mathrm{2D}}=\iint \left\vert U(x,y)\right\vert ^{2}dxdy,  \label{SP2}
\end{equation}%
\begin{equation}
P_{\mathrm{1D}}=\int \left\vert U(x)\right\vert ^{2}dx.  \label{SP1}
\end{equation}

As discussed in the Introduction, the crucially important issues are the ability of the model to produce stable narrow (pixel-like) solitons. As concerns the stability, perturbed solitons solution are introduced in the usual form,
\begin{equation}
E(x,y,z)=[U(x,y)+p(x,y)e^{\lambda z}+q^{\ast }(x,y)e^{\lambda ^{\ast
}z}]e^{ibz},  \label{PERT2D}
\end{equation}%
where $p(x,y)$ and $q(x,y)$ are components of the eigenmode of small perturbations with a stability eigenvalue $\lambda $, the instability taking place if there is, at least, a single eigenvalue with Re$\left( \lambda\right) >0$, and the asterisk ($\ast $) stands for the complex conjugate. In the 1D case, the perturbed solution is sought for as
\begin{equation}
E(x,z)=[U(x)+p(x)e^{\lambda z}+q^{\ast }(x)e^{\lambda ^{\ast }z}]e^{ibz}.
\label{PERT1D}
\end{equation}%
The substitution of the perturbed solution (\ref{PERT2D}) into Eq. (\ref{NLSE}) and linearization with respect to the small perturbations leads to the eigenvalue problem for $\lambda $, represented by the following system of coupled equations:
\begin{equation}
\begin{aligned} i\lambda p=&-\frac{1}{2}\nabla
^{2}p+(b+V)p+\mathrm{g}U(2U^*p+Uq)\\ &+\xi U^{2}U^*(3U^*p+2Uq), \\ i\lambda
q=&+\frac{1}{2}\nabla ^{2}q-(b+V)q-\mathrm{g}U^*(2Uq+U^*p)\\ &-\xi
(U^*)^{2}U(3Uq+2U^*p). \label{LSA} \end{aligned}
\end{equation}

Stationary solutions are found below by means of the squared-operator method \cite{MSOM}, then the eigenvalues of instability growth rate are calculated by the Fourier collocation method \cite{MSOM2}, and the commonly known finite-difference marching scheme is employed to simulate the perturbed propagation of the solitons.

\subsection*{Estimates of the stabilization of the Townes solitons (TSs) by
the lattice potential}

As mentioned above, the simplest stability condition is provided by the VK criterion. In terms of definiton (\ref{U}), it takes the form of
\begin{equation}
dP_{\mathrm{2D,1D}}/db>0.  \label{VKcriterion}
\end{equation}%
Note that the solitons are unstable not only in the case of $dP_{\mathrm{2D,1D}}/db<0,$ but also in the case of the \textit{Townes solitons} (TSs) \textit{viz}., the 1D and 2D ones in the free space with the quintic or cubic self-focusing, respectively, which form degenerate families, whose integral power (norm) does not depend on the propagation constant, $b$, i.e., $dP_{\mathrm{2D,1D}}/db=0$ \cite{Townes,Berge,Salerno}. In particular, the family of the TS solutions of the 1D version of Eq. (\ref{NLSES}), with $\mathrm{g}=0$ and $\xi =1$, is (for all positive values of $b$)
\begin{equation}
U_{\mathrm{TS}}(x;b)=\frac{\left( 3b\right) ^{1/4}}{\sqrt{\cosh \left( 2\sqrt{2b}x\right) }}.  \label{UTS}
\end{equation}
The integral power of this solution indeed does not depend on $b$: $\left( P_{\mathrm{1D}}\right) _{\mathrm{g}=0}=\sqrt{3/2}(\pi /2)$. The initial development of the TS instability is slow (of the power-law type, rather than exponential \cite{Berge}), as it is formally accounted for by vanishing instability growth rates. Therefore, it was possible to experimentally observe weakly unstable 2D TSs in a binary BEC under perturbation-free conditions \cite{Bakkali}.

The VK criterion makes it possible to predict the stabilization of the TSs by the weak lattice potential [presented by the small coefficient $V_{0}\ll 1$ in Eq. (\ref{V1D})] in the limit cases of very broad and very narrow TSs. In the former case, which corresponds to $b\ll 1$ in Eq. (\ref{UTS}), a perturbed solution of the 1D version of Eq. (\ref{NLSES}) with $\mathrm{g}=0$ is looked for as
\begin{equation*}
U(x)=U_{\mathrm{TS}}(x;b)+\delta U_{b\ll 1}(x),
\end{equation*}
with correction $\delta U_{b\ll 1}(x)$ determined by the linearized equation:%
\begin{gather}
-\beta \delta U_{b\ll 1}+\frac{1}{2}\frac{d^{2}}{dx^{2}}\delta U_{b\ll
1}-5U_{\mathrm{TS}}^{4}\left( x;\beta \right) \delta U_{b\ll 1}  \notag \\
=V_{0}~\mathrm{sin}^{2}x\cdot U_{\mathrm{TS}}\left( x;\beta \right) ,
\label{U1}
\end{gather}
with $\beta \equiv b+V_{0}/2$. It is easy to see that, in the limit of $\beta \ll 1$, an approximate solution to Eq. (\ref{U1}) is
\begin{equation}
\delta U_{b\ll 1}(x;\beta )\approx \frac{V_{0}}{4}\cos (2x)\cdot U_{\mathrm{TS}}\left( x;\beta \right) .  \label{U1(x)}
\end{equation}
The respective correction to the integral power $P_{\mathrm{1D}}$ is
\begin{gather}
\delta P_{b\ll 1}(\beta )\approx 2\int_{-\infty }^{+\infty }U_{\mathrm{TS}
}(x;b)\delta U_{b\ll 1}dx  \notag \\
=\sqrt{\frac{3}{2}}\frac{\pi }{4}V_{0}~\mathrm{sech}\left( \frac{\pi }{2
\sqrt{2\beta }}\right) .  \label{deltaP}
\end{gather}
Obviously, this expression produces $d\delta P_{b\ll 1}/db\equiv d\delta P_{b\ll 1}/d\beta >0$, hence the VK criterion (\ref{VKcriterion}) holds for the broad TSs perturbed by the lattice potential, clearly suggesting the stabilization.

In the opposite limit of narrow TSs, which corresponds to $b\gg 1$ in Eq. (\ref{UTS}), it is sufficient to expand potential (\ref{V1D}) around the potential's minimum, $x=0$, which replaces the 1D version of Eq. (\ref{NLSES}) with $\mathrm{g}=0$ by the following equation:
\begin{equation}
-bU=-\frac{1}{2}\frac{d^{2}U}{dx^{2}}+V_{0}x^{2}U-U^{5}.  \label{x^2}
\end{equation}
A simple analysis demonstrates that, in the case of large $b$, the respective correction to the TS solution (\ref{UTS}), produced by the perturbation term $V_{0}x^{2}U$ in Eq. (\ref{x^2}), is
\begin{gather}
\delta U_{b\gg 1}(x;b)\approx \frac{3^{1/4}}{4}\frac{V_{0}b^{-3/4}x^{2}}{
\sqrt{\cosh \left( 2\sqrt{2b}x\right) }}  \notag \\
\times \left[ 1-2\sqrt{2b}x\tanh \left( 2\sqrt{2b}x\right) \right] ,
\label{deltaU}
\end{gather}
cf. Eq. (\ref{U1(x)}). The respective correction to the integral power is
\begin{equation}
\delta P_{b\gg 1}\approx -\frac{\sqrt{3}\pi ^{3}V_{0}}{128\sqrt{2}b^{2}}
\label{deltaP2}
\end{equation}%
[cf. Eq. (\ref{deltaP})], which also satisfies the VK criterion (\ref{VKcriterion}). The fact that the TS family perturbed by the lattice potential may be stable in the limits of the broad and narrow solitons suggests that the entire family may be stable. The full numerical analysis confirms that, indeed, the family of 1D fundamental solitons is almost entirely stable; see Figs. \ref{fig1}(c-f).

\section*{Data availability}
The data supporting the results of this paper are available from the corresponding author upon reasonable request.

\section*{Acknowledgments}

This work is supported by National Natural Science Foundation of China (62205224, 11774068), Guangdong Basic and Applied Basic Research Foundation (2023A1515010865), Guangzhou Science and Technology Plan Project
(2025A04J4068), and Start-up Foundation for Talents of Guangzhou Jiaotong University (K42022076, K42022095).

\section*{Author contributions}

L.Z. has contributed to the idea, the numerical simulations, and written the first draft of the paper. B.A.M. and D.M. carried out the theoretical analysis, discussed the results and revised the manuscript. X.Z. has contributed to the idea, the numerical simulations and revised the manuscript.

\section*{Competing interests}

The authors declare no competing interests.

\end{document}